\documentclass[10pt,a4paper,twoside,final]{article}
\usepackage[latin1]{inputenc}
\usepackage{amsmath}
\usepackage{amsfonts}
\usepackage{amssymb}
\usepackage{url}					
\usepackage[breaklinks]{hyperref}	
\usepackage{graphicx}				

\begin{document}
\bibliographystyle{ijqc}			
\newcommand{\nc}	{\newcommand}
\nc{\rmd}			{\mathrm{d}}
\nc{\rme}			{\mathrm{e}}
\nc{\rmi}			{\mathrm{i}}

\nc{\unitvec}		{\mathbf{e}}
\nc{\unitvecr}		{\unitvec_r}
\nc{\unitvectheta}	{\unitvec_\vartheta}
\nc{\unitvecphi}	{\unitvec_\varphi}
\nc{\polarbasis}	{\left\lbrace\unitvecr,\unitvectheta,\unitvecphi\right\rbrace}

\nc{\deriv}[3]		{\frac{\rmd^{#3} #1}{\rmd #2^{#3}}}
\nc{\funcderiv}[3]	{\frac{\delta^{#3} #1}{\delta #2^{#3}}}
\nc{\partderiv}[3]	{\frac{\partial^{#3} #1}{\partial #2^{#3}}}
\nc{\partder}[3]	{\partial_#2^{#3}#1}

\nc{\approach}[2]	{\mathop{\longrightarrow}_{#1 \to #2}}
\nc{\Ordo}[1]		{\mathrm{O}\left( #1 \right)}
\nc{\abs}[1]		{\left\vert #1 \right\vert}
\nc{\norm}[1]		{\left\Vert #1 \right\Vert}
\nc{\at}[2]			{\left. #1 \right\vert_{#2}}
\nc{\atdiff}[3]		{\left. #1 \right\vert_{#2}^{#3}}

\nc{\trace}[1]		{\Tr \left[ #1 \right]}
\nc{\spur}[1]		{\mathrm{Sp} \left[ #1 \right]}

\nc{\pos}			{\mathbf{r}}
\nc{\Pos}			{\mathbf{R}}
\nc{\posprime}		{\pos^\prime}
\nc{\posdprime}		{\pos^{\prime\prime}}
\nc{\Posprime}		{\Pos^\prime}
\nc{\Posdprime}		{\Pos^{\prime\prime}}
\nc{\dist}[2]		{\abs{\pos_{#1}-\pos_{#2}}}
\nc{\x}				{\mathbf{x}}
\nc{\X}				{\mathbf{X}}
\nc{\Q}				{\mathbf{Q}}

\nc{\thetaphi}		{\vartheta,\varphi}
\nc{\omegaprime}	{\omega^\prime}
\nc{\lm}			{{\ell m}}
\nc{\lambdamu}		{{\lambda\mu}}
\nc{\plambdamu}		{{\lambda^\prime \mu^\prime}}
\nc{\nlm}			{{n \ell m}}
\nc{\nl}			{{n \ell}}
\nc{\sumlambdamu}	{\sum_{\lambda,\mu}}
\nc{\sumplambdamu}	{\sum_{\lambda^\prime, \mu^\prime}}

\nc{\Laplacian}		{\Delta}
\nc{\redmass}		{M}

\nc{\momentum}		{\mathbf{p}}
\nc{\Momentum}		{\mathbf{P}}
\nc{\wavenumber}	{\mathbf{k}}
\nc{\Wavenumber}	{\mathbf{K}}

\nc{\commut}[2]		{[ #1,\, #2 ]}
\nc{\anticommut}[2]	{\{ #1,\, #2 \}}
\nc{\bra}[1]		{\left\langle #1 \right\vert}
\nc{\ket}[1]		{\left\vert #1 \right\rangle}
\nc{\braket}[2]		{\left\langle #1 \vert #2 \right\rangle}
\nc{\matrixel}[3]	{\left\langle #1 \left\vert #2 \right\vert #3 \right\rangle}
\nc{\meanvalue}[2]	{\matrixel{#2}{#1}{#2}}
\nc{\gaunt}[2]		{\braket{#1}{#1 #2}}
\nc{\threej}[6]		{\begin{pmatrix} #1 & #2 & #3 \\ #4 & #5 & #6 \end{pmatrix}}
\nc{\spinsinglet}	{{\uparrow\downarrow}}
\nc{\spintriplet}	{{\uparrow\uparrow}}


\nc{\numden}[1]		{n\left(\x_{#1}\right)}
\nc{\pairden}[2]	{P\left(\x_{#1},\x_{#2}\right)}
\nc{\paircorr}[2]	{g\left(\x_{#1},\x_{#2}\right)}
\nc{\Vee}[1]		{V_{ee}\left(\pos_{#1}\right)}
\nc{\Ne}			{N_e}

\title{Boundary conditions for many-electron systems\thanks{\emph{In memoriam C. Hargitai}}}
\author{P\'eter V. T\'oth}
\maketitle

\begin{abstract}
It is shown that natural boundary conditions for non-relativistic wave functions are of periodic or of homogeneous Robin type. 
Using asymptotic central symmetry of Hamiltonian and  theory of singular differential equations the many-electron wave function is expanded in series both in the vicinity of Coulomb singularities and at infinity. 
Hydrogenic angular dependence of three leading terms of expansion about Coulomb singularities is found.
Exact first- and second-order cusp conditions are obtained demonstrating redundancy of spherical average in Kato's cusp condition.
Our first-order cusp condition exhibits $CP$ symmetry.
Homogeneous Robin boundary conditions are obtained for aperiodic many-electron systems from the expansions.
Use of our explicit boundary conditions improves both speed and accuracy of numerical calculations.
A confluent hypergeometric series defining arbitrarily high order cusp conditions for the spherically averaged Hamiltonian is presented.
\end{abstract}

\section{Introduction\label{sec:intro}}
Boundary conditions play important role in eigenvalue problems of mathematical physics even if they are imposed implicitly.
Explicit use of boundary conditions is crucial to numerical calculations.
The role of regularity and boundary conditions in \emph{existence} of quantum mechanical eigenvalue problems was first recognized by Schr\"odinger \cite{Schrodinger1926} and von Neumann \cite{Neumann1927}.
Boundary and regularity conditions for the Schr\"odinger equation were first studied in detail by Jaff\'e \cite{Jaffe1930, Jaffe1930a} who recognized that both $\psi$ and its first derivatives should be bounded even at singular points of the potential.
McCrea and Newing \cite{McCrea1934} discussed boundary conditions for the $H_2^+$ molecular ion.

In numerical calculations of many-electron systems, explicit boundary conditions should be imposed on the wave function at each singular point of the Hamiltonian, namely, at Coulomb singularities and at infinity 
The exact many-electron wave function satisfies boundary conditions not only at the nuclei but also at electron-electron coalescence points $r_{ij}=0$ $(i \neq j)$.

Asymptotic behaviour of many-electron wave functions was discussed in the framework of Hartree-Fock (HF) approximation by Handy, Marron and Silverstone \cite{Handy1969} and in the general case by Katriel and Davidson \cite{Katriel1980} and by Patil \cite{Patil1989, Patil1990}.
Vitanov and Panev \cite{Vitanov1992} extended the investigation to excited atomic states.
The long-range behaviour of density was studied by Tal \cite{Tal1978} and by Levy, Perdew and Sahni \cite{Levy1984}.
Ernzerhof, Burke and Perdew \cite{Ernzerhof1996} discussed the long-range behaviour of ground-state wave functions, one-electron density matrices, and pair densities including their angular dependence.

The study of many-electron wave functions about Coulomb singularities dates back to L\"owdin \cite{Lowdin1954} who first recognized their cusp-like behaviour.
He obtained, by analysing tabulated atomic HF data, hydrogenic $r$-dependence $R_\nl(r) = r^\ell f_\nl(r) \propto r^\ell [1 - Zr/(\ell+1) + \ldots]$ of radial wave functions in the vicinity of  nucleus resulting in \emph{cusp condition} 
\begin{equation} \label{eq:lowdin} 
 \at{\frac{f^\prime_\nl}{f_\nl}}{r=0} = -\frac{Z}{\ell+1}.
\end{equation}
Similar conditions were obtained both for electron-nucleus and electron-electron coalescences in the ground state of the $He$ atom by Roothaan and Weiss \cite{Roothaan1960} by analysing singularities of Hamiltonian which were used as constraints on trial functions in their variational calculation in order to avoid divergence of local energy $E_{loc}\equiv\Psi^{-1}\hat{H}\Psi$ at coalescence points $r_1=0$, $r_2=0$ and $r_{12}=0$.
They have introduced the \emph{reduced mass} of coalescent particles into the cusp condition
(see quotation from p. 196 of Ref. \cite{Roothaan1960}):
\begin{quote}
Each cusp refers to a pair of Coulombic particles; the sign of the cusp is positive for a repulsion, negative for an attraction, and the magnitude is equal to the product of the two charges and the reduced mass of the two particles.
\end{quote}
Roothaan, Sachs and Weiss \cite{Roothaan1960a} used above cusp condition in variational calculation of light atoms and ions (up to $Z=10$).
The first variational molecular calculation using the cusp constraint (for $H_2$) was performed by Ko\l{}os and Roothaan \cite{Kolos1960}.

The near-nucleus behaviour of an arbitrary many-electron eigenfunction $\Psi=\Psi(\pos\vert\pos_2,\ldots,\pos_N)$ was given by Kato's \cite{Kato1957} theorem $IIb$
\begin{equation}	\label{eq:kato} 
 \at{\frac{\partial \overline{\Psi}}{\partial r}}{r=0} = -Z \Psi(r=0),
\end{equation}
where $Z$ denotes nuclear charge of the atom located at the origin and overline symbol stands for the spherical average.
Steiner \cite{Steiner1963} added a corollary to the theorem which relates the charge density and its derivative at the nucleus of an atom in a similar manner.
M. Hoffmann-Ostenhof and Seiler \cite{Hoffmann-Ostenhof1981} generalized Kato's result to multiple coalescences; their fixed-nucleus approximation was removed by Johnson \cite{Johnson1981} using Jacobi coordinates.
Validity of Eq. (\ref{eq:kato}) is limited to the $s$ states since wave function $\psi \propto r^\ell$ has a root of multiplicity $\ell$ at the origin.
Unfortunately, Eq. (\ref{eq:kato}) and its generalizations cannot be used as exact boundary conditions due to the spherical average.
It will be shown in section \ref{sec:cusp-bc} that the \emph{spherical average is redundant in Kato's cusp condition} (\ref{eq:kato}) which was already indicated for above mentioned few-electron systems by Roothaan and his co-workers \cite{Roothaan1960, Roothaan1960a, Kolos1960}.

Results of our thorough and rigorous analysis differ from that of three highly-cited papers \cite{Pack1966, Bingel1963, Bingel1967} but are in accordance with that of L\"owdin \cite{Lowdin1954} and of Roothaan and his co-workers \cite{Roothaan1960, Roothaan1960a, Kolos1960}.
Pack and Brown \cite{Pack1966} added the phrase "all $l, m$" to L\"owdin's cusp condition (\ref{eq:lowdin}) by solving many-electron Schr\"odinger equation in the vicinity of coalescence points which is not justified by our more rigorous solution.
Bingel \cite{Bingel1963, Bingel1967} introduced an additional linear term proportional to the cosine of an unknown angle in the series expansion of wave function of which spherical average is always zero.
In our rigorously derived expansion, all three leading terms exhibit the same hydrogenic angular dependence $Y_\lm(\vartheta, \varphi)$ as a result of isotropy of three leading terms of many-electron Hamiltonian in the vicinity of Coulomb singularities.
Perhaps some recent research (e.g. Refs. \cite{Pan2003, March2000}) relying on these old results should be revised.

As a result of spherical average introduced by Kato the higher-order cusp conditions were derived only for the spherically averaged Hamiltonian by Rassolov and Chipman \cite{Rassolov1996} which were generalized to the excited states by Nagy and Sen \cite{Nagy2000, Nagy2000a, Nagy2001}.
The near-nucleus anisotropy of Hamiltonian was first taken into consideration by Qian and Sahni \cite{Qian2007}, nevertheless omitting term $\Ordo{r^0}$. 
We will show in sections \ref{subsec:Psi-0} and \ref{subsec:cusp} that first and second order cusp conditions are exact without spherical averaging and only that of third and of higher orders need spherical average.
In addition, we present a confluent hypergeometric series in section \ref{subsec:Psi-0-avg} defining arbitrarily high order cusp conditions for the spherically averaged Hamiltonian.

Fournais, M. and T. Hoffmann-Ostenhof and S\o{}rensen \cite{Fournais2009} recently decomposed the many-electron wave function as $\psi(\mathbf{x}) = \psi^{(1)}(\mathbf{x}) + \abs{x_1} \psi^{(2)}(\mathbf{x})$ in the neighbourhood of Coulomb singularities using inconvenient Cartesian coordinates $\mathbf{x} = \left(x_1,\ldots,x_N\right) \in \mathbb{R}^{3N}$ inhibiting the recognition of asymptotic central symmetry of Hamiltonian. 
In addition, their component functions $\psi^{(1)}$ and $\psi^{(2)}$ are real-valued.
Our decomposition (\ref{eq:Psi-ser-0}) is slightly different due to the isotropy of singularity taken into consideration: the first term is an eigenfunction of the spherically averaged Hamiltonian, the second term reflects the anisotropy of molecular or crystalline potential.

Since Kato's rigorously proved spherically averaged cusp condition (\ref{eq:kato}) contradicts Eq. (\ref{eq:lowdin}) which was successfully used in few-electron calculations by Roothaan and his co-workers, a different but similarly rigorous treatment is needed to decide between the two forms.
The primary aim of this paper is to derive boundary conditions by solving many-electron Schr\"odinger equation both at large distances and in the neighbourhood of Coulomb singularities using a well-tried formalism of  mathematical physics, the theory of singular differential equations.
The key idea of this paper is that due to \emph{asymptotic isotropy} \footnote{Term \emph{asymptotic equality} is used in this paper in the sense of $\lim_{x \to x_0} \left[f(x)/g(x)\right] = 1$. Isotropic and anisotropic terms of expansions are distinguished throughout this paper so a boldface argument of \emph{ordo} symbol $\Ordo{\pos^n}$ indicates the anisotropy of omitted terms of expansions.} 
of many-electron Hamiltonian both at Coulomb singularities and as $r \to \infty$ the angular momentum asymptotically commutes with the Hamiltonian leading to asymptotically hydrogen-like wave functions with definite values of quantum numbers $\ell$ and $m$.

The outline of the paper is as follows.
In section \ref{sec:bc-qm}, the allowed boundary conditions of non-relativistic one-particle quantum mechanics are studied.
In section \ref{subsec:Isotropy}, the isotropy of singularities of many-electron Hamiltonian is discussed.
In section \ref{subsec:Psi-inf}, the many-electron wave function is expanded at large distances in terms of irregular solid spherical harmonics (\ref{eq:Psi-ser-inf}). 
At the beginning of section \ref{subsec:Psi-0}, the many-electron Hamiltonian (\ref{eq:hamiltonian-0}) is studied in the vicinity of Coulomb singularities, where explicit form (\ref{eq:W-r}) of potential contribution of non-coalescent particles is obtained.
Then it is shown that due to isotropy of leading terms of Hamiltonian, all three leading terms of many-electron wave function exhibit hydrogenic angular dependence (\ref{eq:Psi-ser-0}), therefore first and second order cusp conditions (\ref{eq:u-cusp1}, \ref{eq:u-cusp2}) are exact without spherical averaging.
The section ends with the decomposition of many-electron wave function as the sum of two functions, where the first term is an eigenfunction of the spherically averaged Hamiltonian, the second term reflects the anisotropy of molecular or crystalline potential.
In section \ref{subsec:Psi-0-avg}, eigenfunctions of the spherically averaged Hamiltonian are discussed leading to recurrence relations (\ref{eq:recurrence-1}, \ref{eq:recurrence-k}) defining cusp conditions of arbitrarily high orders for this special case.
In section \ref{subsec:cusp}, exact first and second order cusp conditions (\ref{eq:Psi-cusp-nuc}, \ref{eq:Psi-cusp-el-sing}, \ref{eq:Psi-cusp-el-trip}, \ref{eq:Psi-cusp2}) are presented demonstrating the redundancy of spherical average in Kato's cusp condition (\ref{eq:kato}).
As a result of asymptotic hydrogenic behaviour of many-electron wave function both at Coulomb singularities and as $r \to \infty$ we obtain boundary conditions of homogeneous Robin type (\ref{eq:bc-Psi-inf}, \ref{eq:bc-Psi-0}) similarly to Eqs. (\ref{eq:bc-inf}) and (\ref{eq:bc-nuc}) derived in section \ref{sec:bc-qm} for the $H$ atom.
In section \ref{sec:consequences}, physical and numerical consequences of our results are discussed.

\section{Boundary conditions in one-particle quantum mechanics \label{sec:bc-qm} }
The problem of hydrogen atom was solved by Schr\"odinger in his historically famous paper  \cite{Schrodinger1926} without imposing explicit boundary conditions on the wave function.
In the first version of manuscript, the requirement of stationarity of current flux was used as a constraint on the variational problem which was changed to the weaker normalization condition by an addendum (cf. equations 6 and 24 of Ref. \cite{Schrodinger1926}).
Latter form is more conventional mathematically since it is compatible with Sturm-Liouville theory of eigenvalue equations, where $\int \abs{\psi}^2 dv$ represents the denominator of Rayleigh quotient.
Von Neumann has concluded that normalization condition for the wave function and requirement of self-adjointness of the Hamiltonian is equivalent to imposing both boundary and regularity conditions on the wave function \cite{Neumann1927}.

In fact, these conditions are too weak to enforce unique regular solutions of Schr\"{o}dinger equation.
The normalization condition does not exclude irregular particular solution $\psi \propto r^{-\ell-1}$ for $s$ states of the Coulomb problem \cite{Ballentine1998} hence it is excluded by hand both in Schr\"odinger's paper \cite{Schrodinger1926} and in the textbooks.
The requirement of self-adjointness does not lead to a unique eigenvalue problem since a little-known theorem \cite{Korn1968, Ince1944} of Sturm-Liouville theory of differential equations states that \emph{any} of following two types of boundary conditions are consistent with self-adjointness of the Liouville operator:
\begin{enumerate}
 \item \emph{periodic} boundary conditions
  \begin{subequations}
  \begin{eqnarray} 
  \psi(a)        - \psi(b)        &=& 0,  \label{eq:bc-periodic} \\
  \psi^\prime(a) - \psi^\prime(b) &=& 0,  \label{eq:bc-periodic2}  
  \end{eqnarray}
  \end{subequations}
 \item \emph{homogeneous Robin} boundary conditions
  \begin{subequations}
  \begin{eqnarray}   
  \alpha_1 \psi^\prime(a) + \beta_1 \psi(a) &=& 0,  \label{eq:bc-robin} \\
  \alpha_2 \psi^\prime(b) + \beta_2 \psi(b) &=& 0,  \label{eq:bc-robin2} 
  \end{eqnarray}
  \end{subequations}
\end{enumerate}
where $\alpha$'s and $\beta$'s are real constants, $a$ and $b$ denote endpoints of the interval \footnote{The natural boundary condition for the momentum operator $-\rmi\deriv{}{x}{}$, discussed by von Neumman \cite{Neumann1927} as an example, is simply (\ref{eq:bc-periodic}). Widely used substitution $\psi=u(r)/r$ eliminates not only the first derivative from the Laplacian but also the periodic boundary conditions since $r$ equals the square root of coefficient $p=r^2$ of Liouville operator $\deriv{}{r}{}\left[p(r) \deriv{}{r}{}\right]+q(r)$, i.e. only one branch of $\pm\sqrt{p}$ is used which illustrates the importance of single-valuedness of coefficients in the theory of differential equations. In $d$-dimensional hyperspherical coordinates, the substitution eliminating the first derivative is of the form $\psi=u(r)/r^{(d-1)/2}$, where the square root appears explicitly.}.
The theorem can be generalized to partial Sturm-Liouville equations by taking function values and normal derivatives over hypersurfaces of the domain (the proof is based on the definition of self-adjointness and the use of Green's theorem).

Equations (\ref{eq:bc-periodic}, \ref{eq:bc-periodic2}) are known as Born - von K\'arm\'an \cite{Born1912, Born1913, Born1913a} or Bloch \cite{Bloch1928} boundary conditions of solid state physics.
In view of above and Bloch's theorem we can state that \emph{eigenfunctions of aperiodic systems satisfy homogeneous Robin boundary conditions}.
Homogeneous Dirichlet and Neumann boundary conditions for model problems of the textbooks are special cases of (\ref{eq:bc-robin}, \ref{eq:bc-robin2}).
Boundary conditions (\ref{eq:bc-robin}, \ref{eq:bc-robin2}) can be divided by arbitrary constants so coefficients $\alpha_i/\sqrt{\alpha_i^2+\beta_i^2} \equiv \sin\gamma_i$ and $\beta_i/\sqrt{\alpha_i^2+\beta_i^2} \equiv \cos\gamma_i$ define angles $\gamma_1$ and $\gamma_2$ representing the boundaries.
Random coefficients for amorphous materials, in the Wannier \cite{Wannier1937} representation, may be interpreted as random walk of these "phase points" around the unit circle which leads to a band structure similarly to the periodic boundary conditions.

As an example of Eqs. (\ref{eq:bc-robin}, \ref{eq:bc-robin2}) let us recover hidden boundary conditions for a non-relativistic $H$-like ion with nuclear charge $Z$ using known properties of hydrogenic bound-state wave functions.
The normalization condition guarantees a part of boundary conditions, namely, vanishing at infinity.
Asymptotic exponential decay of wave function $\psi=\psi(\pos)$ is described by limit of logarithmic derivative
\begin{equation}   \label{eq:bc-inf}
 \lim_{r\to\infty} \frac{1}{\psi} \frac{\partial\psi}{\partial r}=-\sqrt{-2E}, 
\end{equation}
where, and throughout this paper, atomic units ($\hbar=e=m_e=4\pi\varepsilon_0=1$) 
are used.
The normalization condition results in a mild singularity (hyperconical cusp) of the wave function: whereas it is continuous everywhere, its directional derivatives are bounded but discontinuous at the Coulomb singularity. 
Since eigenfunctions of the central field problem are separable as $\psi=R_\nl(r)Y_\lm(\thetaphi)$, their directional logarithmic derivatives are of the form
\begin{equation*}
   \frac{\unitvec\cdot\nabla\psi}{\psi} 
 = \frac{\unitvec\cdot\unitvecr}{R_\nl} \deriv{R_\nl}{\,r}{}
 + \frac{\unitvec\cdot\unitvectheta}{Y_\lm\,r} 
   \frac{\partial Y_\lm}{\partial\vartheta}
 + \frac{\unitvec\cdot\unitvecphi}{Y_\lm\,r\sin\vartheta} 
   \frac{\partial Y_\lm}{\partial\varphi},  
\end{equation*}
where $\unitvec$ stands for unit vector of the selected direction, $\polarbasis$ is the basis of the spherical polar coordinate system.
In radial directions defined by $\unitvec\cdot\unitvecr = \pm 1$ and $\unitvec\cdot\unitvectheta = \unitvec\cdot\unitvecphi = 0 $, above expression reduces to
\begin{equation*}
   \frac{\unitvec\cdot\nabla\psi}{\psi} 
 = \frac{\pm 1}{R_\nl} \deriv{R_\nl}{\,r}{}
 = \frac{\pm 1}{\psi} \frac{\partial \psi}{\partial r}. 
\end{equation*}
Since radial wave function $R_\nl = r^\ell u_\nl(r)$ of the central field problem has a root of multiplicity $\ell$ at the origin the l'Hospital rule should be applied $\ell$ times in order to obtain a definite limit
\begin{equation*}
   \lim_{r \to 0} \frac{1}{R_\nl} \deriv{R_\nl}{\,r}{}
 = \lim_{r \to 0} \frac{R_\nl^{(\ell+1)}}{R_\nl^{(\ell)}}
 = (\ell+1) \frac{u^\prime_\nl(0)}{u_\nl(0)} = -Z, 
\end{equation*}
where differentiation rules (\ref{eq:der-l}, \ref{eq:der-l+1}) and explicit form of the hydrogenic radial wave function 
\footnote{The power series expansion of the hydrogenic radial wave function about the nucleus is of the form $R_\nl = C_\nl\, r^\ell \left[ 1 - \frac{Zr}{\ell+1} + \frac{2n^2+\ell+1}{2(\ell+1)(2\ell+3)} \left(\frac{Zr}{n}\right)^2 + \ldots\right]$.} 
are used.
The discontinuity of the radial logarithmic derivative at the nucleus is then characterized by \begin{equation}	\label{eq:bc-nuc} 
 \lim_{\mathbf{r} \to \pm 0} \frac{1}{\psi} \frac{\partial \psi}{\partial r}
 = \lim_{\mathbf{r} \to \pm 0} \frac{\partial_r^{\ell+1}\psi}{\partial_r^\ell \psi}
 = \mp Z,
\end{equation}
where
\begin{eqnarray*}
 \mathbf{r} \to +0 &\equiv& (r \to 0, \,\vartheta,     \,\varphi),		\\
 \mathbf{r} \to -0 &\equiv& (r \to 0, \,\pi-\vartheta, \,\pi+\varphi).
\end{eqnarray*}
We emphasize that all steps of above derivation, except the last one, rely on isotropy of singularity of Hamiltonian at $r=0$.
It is interesting to observe the $CP$ invariance of cusp relation (\ref{eq:bc-nuc}).
Pair of (\ref{eq:bc-inf}) and (\ref{eq:bc-nuc}) obviously represent homogeneous Robin boundary conditions of the form (\ref{eq:bc-robin}, \ref{eq:bc-robin2}).

Similar boundary conditions will be obtained for many-electron wave functions in section \ref{sec:cusp-bc} as a result of asymptotic central symmetry of many-electron Hamiltonian both in the vicinity of nuclei and at large distances.

\section{Behaviour of many-electron wave function at singular points of Hamiltonian \label{sec:wave-fun}}
Let us consider non-relativistic Hamiltonian describing $N$ particles interacting with each other by Coulomb potentials 
\begin{equation}	\label{eq:hamiltonian}
 \hat{H} = -\frac{1}{2}\sum_{i=1}^N \frac{\Laplacian_i}{m_i}
         + \sum_{i=1}^{N-1} \sum_{j=i+1}^N \frac{q_i q_j}{r_{ij}} ,
\end{equation}
where $m_i$ and $q_i$ denote mass and charge of the $i$-th particle, respectively ($m_i=1$ and $q_i=-1$ for electrons and $m_i=1836A_\nu$, $q_i=Z_\nu$, $\nu=1,2,\ldots<N$ for nuclei).
Electrons and nuclei will be distinguished only in the final results.
Spin coordinates are omitted for simplicity.
Many-particle wave functions $\Psi=\Psi\left( \pos_1, \pos_2, \ldots , \pos_N \right)$ satisfy stationary-state Schr\"odinger equation
\begin{equation}	\label{eq:schrodinger}
 \hat{H}\Psi = E \Psi 
\end{equation}
which is singular both at coalescence points $\pos=\pos_i$ and as $r\to\infty$.
Singularities of the Coulomb potential are removable and singularity of the kinetic energy at infinity is essential.
The singularities and other leading terms of the Hamiltonian are isotropic in both limiting cases.
In the vicinity of coalescence points, the singular potential contribution of coalescent particles $\Ordo{r^{-1}}$ and bounded leading term of potential due to remaining non-coalescent particles $\Ordo{r^0}$ are isotropic.
At infinity, only Coulombic monopole term $\Ordo{r^{-1}}$ is isotropic, whereas multipole  terms $\Ordo{\pos^{-2}}$ are anisotropic.

\subsection{Isotropy of singularities \label{subsec:Isotropy} }
Particular solutions of a linear ordinary differential equation in the neighbourhood of an isolated singular point $x_0=0$ are of product form $y=f(x) g(x)$, where $f(x)$ ensures the correct behaviour of $y$ at the singularity and $g(x)$ is a single-valued analytic function (or has at most a logarithmic singularity) which is non-zero at the singular point. Function $f(x)$ is typically power, exponential or Gaussian function.
For \emph{isotropic singularities} of a linear partial differential equation, the solution has the form $y=f(r) g(\pos)$ with $\pos\in\mathbb{R}^n$ and $r=\abs{\pos}$, where $f(r)$ is responsible for correct behaviour of $y$ about the singular point and $g(\pos)$ (reflecting anisotropy of the coefficient functions) is non-zero at the singular point.
(One may expand it in 3 dimensions in terms of regular $r^\lambda Y_\lambdamu(\thetaphi)$ or irregular $r^{-\lambda-1} Y_\lambdamu(\thetaphi)$ solid spherical harmonics.)
It will be shown in this section that due to isotropy of singularities of Hamiltonian (\ref{eq:hamiltonian}), the wave function about singular points has the following limiting forms
\begin{eqnarray*}
 && \Psi \approach{\pos}{\pos_i=0} r^{\ell_i} u_i(\pos), 
    \quad u_i(0) \neq 0, \; i=1,2,\ldots,N,	\\
 && \Psi \approach{r}{\infty} e^{\alpha r} r^\beta v(\pos), 
    \quad \lim_{r\to\infty}v(\pos)\neq 0, \; \alpha<0.
\end{eqnarray*}

In many-body systems, in contrast to the central-field problem, only energy $E$ is conserved throughout the configuration space, square $\mathbf{L}^2$ and projection $L_z$ of angular momentum are conserved only at the singularities
\begin{subequations}
\begin{eqnarray}
 \commut{\hat{H}}{\hat{\mathbf{L}}^2} \approach{\pos}{\pos_i} 0,		\quad
 && \commut{\hat{H}}{\hat{L}_z} \approach{\pos}{\pos_i} 0, \label{eq:commut-0}	\\
 \commut{\hat{H}}{\hat{\mathbf{L}}^2} \approach{r}{\infty} 0,			\quad 
 && \commut{\hat{H}}{\hat{L}_z} \approach{r}{\infty} 0,    \label{eq:commut-inf} 
\end{eqnarray}
\end{subequations}
where Hamiltonian (\ref{eq:hamiltonian}) is rotationally invariant.
In the neighbourhood of singular points angular momentum quantum number $\ell$ and magnetic quantum number $m$ have definite values.
Eigenfunctions of many-electron Hamiltonian approach eigenfunctions of angular momentum when approaching singularities of the Hamiltonian
leading to \emph{asymptotic hydrogenic angular dependence}
\begin{subequations}
\begin{eqnarray}
 \label{eq:Psi-hyd-0} 
 &&\Psi \approach{\pos}{\pos_i=0} R_i(r) Y_{\ell_i m_i}(\thetaphi), \\
 \label{eq:Psi-hyd-inf} 
 &&\Psi \approach{r}{\infty} R_\infty (r) Y_{00}
\end{eqnarray}
\end{subequations}
 of the wave function, where the power and exponential functions reflecting the singularities are included in the radial functions.
In other words, molecular symmetries manifest only at molecular distances, where many-body Hamiltonian does not commute with angular momentum.

Due to asymptotic $H$-like behaviour of many-electron wave functions the equations of this section and their solution methods are similar to the mathematically correct original treatment of $H$ atom  by Schr\"odinger and Weyl (see acknowledgement in footnote on p. 363 of Ref.  \cite{Schrodinger1926})
based on theory of singular differential equations
which differs from the simplistic textbook-derivations.
The Coulomb singularities are treated using Fuchs' theorem exactly the same way as in Ref. \cite{Schrodinger1926}. The singularity at $\infty$ is treated using Hamburger's theorem which is shorter than Schr{\"o}dinger's solution based on the Laplace transform. 
Of course, our equations lead to wave functions of the $H$ atom in the special case of two Coulombic particles.

\subsection{Asymptotic behaviour at infinity \label{subsec:Psi-inf} }
Let us consider an electron, say particle 1, separated from the rest of the system ($m_1 = 1$, $q_1 = -1$ and $r_1 > r_2, \ldots, r_N$). 
Let us introduce reduced mass $\redmass^\prime$ and center of mass $\Posprime$ of the whole system, center of mass $\Posdprime$ of particles except the electron located at $\pos_1$ and their separation $\pos$:
\begin{eqnarray*}
 &&\frac{1}{\redmass^\prime}	\equiv \sum_{i=1}^N \frac{1}{m_i},		\quad
 \Posprime  \equiv \frac{\sum_{i=1}^N m_i \pos_i}{\sum_{i=1}^N m_i},	\quad
 \Posdprime \equiv \frac{\sum_{i=2}^N m_i \pos_i}{\sum_{i=2}^N m_i},	\\
 &&\pos \equiv \pos_1 - \Posdprime \equiv  (r,\thetaphi) \equiv (r,\omega).
\end{eqnarray*}
Using Laplace expansion for $r > r_i$
\begin{equation*}
 \frac{1}{\abs{\pos-\pos_i}} 
 = \sum_{\lambda=0}^\infty \sum_{\mu=-\lambda}^\lambda
   \frac{4\pi}{2\lambda+1} \frac{r_i^\lambda}{r^{\lambda+1}}
   Y_\lambdamu^\ast(\omega) Y_\lambdamu(\omega_i)  
\end{equation*}
the potential energy of the system can be expressed at large distances as
\begin{eqnarray*}
 U(\pos) &=& \sumlambdamu \frac{4\pi q_1}{2\lambda+1}
  \frac{Y_\lambdamu^\ast(\omega)}{r^{\lambda+1}}
  \sum_{i=2}^N q_i r_i^\lambda Y_\lambdamu(\omega_i)		\\
 &=& \frac{q_1}{r} \sum_{i=2}^N q_i + \Ordo{\pos^{-2}} 
 = -\frac{Q+1}{r} + \Ordo{\pos^{-2}},
\end{eqnarray*}
where $Q \equiv \sum_{i=1}^N q_i$
by noting that $Q=0$ for neutral systems.

Due to asymptotic isotropy (\ref{eq:commut-inf}) of Hamiltonian the many-particle Schr\"odinger equation is asymptotically separable (\ref{eq:Psi-hyd-inf}) in terms of spherical polar coordinates as $r \to \infty$.
Asymptotic radial wave function $R=R\left(r\vert\pos_2,\ldots,\pos_N\right)$ satisfies differential equation \footnote{Centrifugal kinetic energy term is omitted since it has the same order as omitted anisotropic dipole term of the potential energy.}
\begin{equation}	\label{eq:rad-inf} 
 \left[ -\frac{\Laplacian_r}{2\redmass^\prime} - \frac{Q+1}{r} - E + \Ordo{\pos^{-2}} \right] R \approach{r}{\infty} 0,
\end{equation}
which is not an eigenvalue equation and $E$ is eigenvalue of (\ref{eq:schrodinger}).
The equation has an isolated \emph{essential} singularity at infinity since transformation of variable $z \equiv 1/r$ leads to second order differential equation
\[ z^4 R^{\prime\prime}(z) + 2\redmass^\prime \left[ E + (Q+1)z + \Ordo{\mathbf{z}^2} \right] R(z) \approach{z}{0} 0  \]
with an isolated fourth order pole at the origin. The pole being independent of potential is a consequence of the Laplacian.
Radial equation (\ref{eq:rad-inf}) is of Hamburger type \cite{Smirnov1974, Smirnov1964, Ince1944}
\begin{equation*}
 \deriv{R}{\,r}{2}
 + \left( a_0+\frac{a_1}{r}+\frac{a_2}{r^2}+\ldots \right) \deriv{R}{\,r}{}
 + \left( b_0+\frac{b_1}{r}+\frac{b_2}{r^2}+\ldots \right) R = 0  
\end{equation*}
with
\begin{equation*}
 a_1 = 2, \quad a_0 = a_2 = a_3 = \ldots = 0,	\quad
 b_0 = 2\redmass^\prime E, \quad b_1 = 2\redmass^\prime (Q+1),
\end{equation*}
$b_2$ and higher order coefficients are anisotropic.
Due to essential singularity of the equation the solution vanishes transcendentally as $r \to \infty$.
Seeking the solution in the form
\[ R = \rme^{\alpha r} u(r), \quad \lim_{r\to\infty} u(r) \neq 0 \]
we obtain
\begin{equation*}
 \deriv{u}{r}{2}	 
 + 2\left( \alpha + \frac{1}{r}\right) \deriv{u}{r}{}
 + \left[ \alpha^2 + b_0 + \frac{2\alpha+b_1}{r} + \Ordo{\pos^{-2}} \right] u 
 \approach{r}{\infty} 0. 
\end{equation*}
By equating leading term of coefficient of $u(r)$ to zero we obtain \emph{indicial equation} of which roots are 
\[ \alpha = \pm \sqrt{-b_0}, \]
where only the minus sign leads to a bounded solution.
Non-essential singularity of the latter differential equation can be removed by substitution
\[ u = r^\beta v(r), \quad \lim_{r\to\infty} v(r) \neq 0 \]
yielding
\begin{equation*}
 \deriv{v}{r}{2} 
 + 2 \left( \alpha + \frac{\beta+1}{r} \right) \deriv{v}{r}{}
 + \left[ \frac{2\alpha(\beta+1)+b_1}{r} + \Ordo{\pos^{-2}} \right] v 
 \approach{r}{\infty} 0.
\end{equation*}
By equating leading term of coefficient of $v(r)$ to zero we obtain indicial equation with only root 
\[ \beta = -\frac{b_1}{2\alpha} - 1 = \frac{b_1}{2\sqrt{-b_0}} - 1. \]
In case of a central-symmetric problem the solution of latter differential equation would be of the form
\[ v = v_0 + v_1 r^{-1} + v_2 r^{-2} + \ldots, \quad (v_0 \neq 0) \]
but since $b_2$ is anisotropic we restrict our solution to $v_0$.
The radial wave function at large distances is then
\begin{equation}	\label{eq:R-inf} 
 R \approach{r}{\infty} 
 \rme^{-\sqrt{-2\redmass^\prime E}\,r}
 r^{\frac{\redmass^\prime (Q+1)}{\sqrt{-2\redmass^\prime E}}-1}
 \left[ v_0 + \Ordo{\pos^{-1}} \right] .
\end{equation}
This behaviour is in accordance with results of Katriel and Davidson \cite{Katriel1980} who derived it less rigorously in two different ways.
Since above function completely characterizes the singularity of Hamiltonian at $\infty$ the many-electron wave function can be expanded in terms of irregular solid spherical harmonics as
\begin{equation}	\label{eq:Psi-ser-inf} 
 \Psi = \rme^{-\sqrt{-2\redmass^\prime E}\,r} r^{\frac{\redmass^\prime (Q+1)}{\sqrt{-2\redmass^\prime E}}}
 \sumlambdamu \frac{v_\lambdamu\left(r^{-1}\vert\pos_2,\ldots,\pos_N\right)}{r^{\lambda+1}} Y_\lambdamu(\thetaphi),
\end{equation}
where $v_{00} = \sqrt{4\pi} \, v_0 =$ constant and coefficients $v_\lambdamu\left(r^{-1}\vert\pos_2,\ldots,\pos_N\right)$ are single-valued analytic functions of $r^{-1}$. 
Molecular or crystalline symmetries are reflected by relations between coefficient functions $v_\lambdamu\left(r^{-1}\vert\pos_2,\ldots\pos_N\right)$ for $\lambda>0$.

\subsection{Local behaviour at Coulomb singularities \label{subsec:Psi-0}}
\begin{figure} 
\includegraphics[width=4.5in]{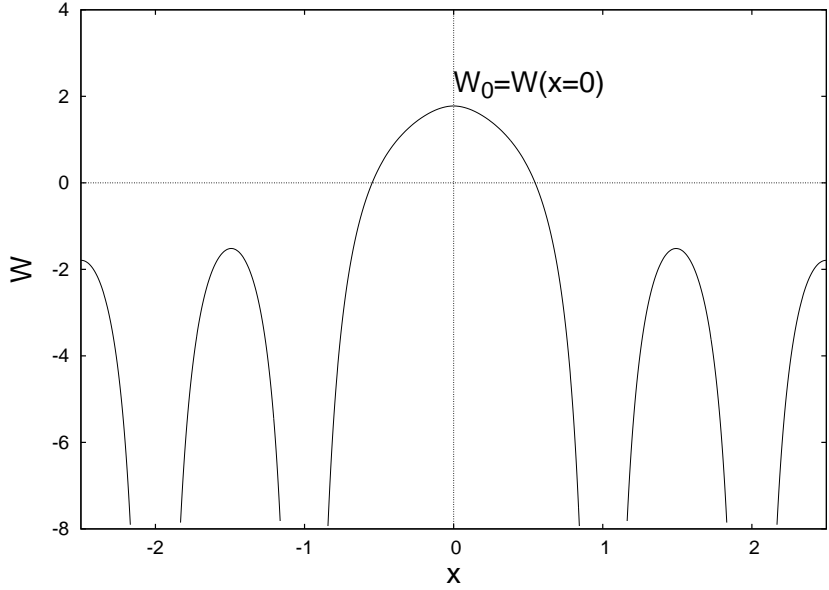}
\caption{\label{fig:W-chain}
Illustrative example of a linear chain of homo-nuclear atoms showing the behaviour of potential term $W$ (\ref{eq:W-r}) of Hamiltonian (\ref{eq:hamiltonian-0}) in the vicinity of an electron-nucleus coalescence point $x=0$.}
\end{figure}

Let us focus now our attention on coalescence of any two particles, say $1$ and $2$, while keeping remaining particles separated from them: $r_1,\,r_2 < r_3, \ldots, r_N$ \footnote{Other authors e.g. of Refs. \cite{Pack1966, Rassolov1996} assume well separated particles $r_1,\,r_2 \ll r_3, \ldots, r_N$ which is unnecessary.}.
In order to explore symmetry properties of the Hamiltonian it is convenient to use Jacobi coordinates by introducing reduced mass $\redmass$, center of mass $\Pos$ and separation $\pos$ of these two particles:
\begin{eqnarray*}
 \redmass  \equiv  \frac{m_1 m_2}{m_1 + m_2},	\quad
 \Pos	   \equiv  \frac{m_1\pos_1 + m_2\pos_2}{m_1+m_2}, \quad \\
 \pos	   \equiv  \pos_1 - \pos_2 \equiv (r,\thetaphi) \equiv (r,\omega).	 
\end{eqnarray*}
Many-body Hamiltonian (\ref{eq:hamiltonian}) can be partitioned as
\begin{equation}	\label{eq:hamiltonian-0} 
 \hat{H} = - \frac{\Laplacian}{2\redmass} + \frac{q_1 q_2}{r} + \hat{W} + \hat{G},
\end{equation}
where
\begin{eqnarray*}	
 \hat{W} & \equiv &   \sum_{i=3}^N \left( \frac{q_1}{r_{1i}} 
                    + \frac{q_2}{r_{2i}} \right) q_i,	\\
 \hat{G} & \equiv & - \frac{\Laplacian_{\Pos}}{2(m_1 + m_2)}
                    - \sum_{i=3}^N \frac{\Laplacian_i}{2m_i}
		            + \sum_{i=3}^{N-1} \sum_{j=i+1}^N \frac{q_i q_j}{r_{ij}} .
\end{eqnarray*}
Use of Laplace expansion for $r < r^\prime$
\begin{equation*}
 \frac{1}{\abs{\pos-\posprime}} = \sum_{\lambda=0}^\infty \sum_{\mu=-\lambda}^\lambda
 \frac{4\pi}{2\lambda+1} \frac{r^\lambda}{r^{\prime\lambda+1}}
 Y_\lambdamu^\ast(\omega) Y_\lambdamu(\omega^\prime) 
\end{equation*}
yields
\begin{eqnarray*}
 W = &&\sumlambdamu \frac{4\pi w_\lambdamu}{2\lambda+1}
     \left[ q_1 r_1^\lambda Y_\lambdamu^\ast(\omega_1) 
     + q_2 r_2^\lambda Y_\lambdamu^\ast(\omega_2) \right],	\\
 &&w_\lambdamu \equiv w_\lambdamu \left(\pos_3,\ldots,\pos_N\right) 
   \equiv \sum_{i=3}^N \frac{q_i}{r_i^{\lambda+1}} Y_\lambdamu(\omega_i).
\end{eqnarray*}
Expressing $\pos_1$ and $\pos_2$ with Jacobi coordinates $\pos$ and $\Pos$, putting origin of the coordinate system to center of mass $\Pos$, using inversion property
\begin{equation*}
 Y_\lambdamu(\pi-\vartheta, \,\pi+\varphi) = (-1)^\lambda\, 
 Y_\lambdamu(\thetaphi) 
\end{equation*}
and addition theorem
\begin{eqnarray*}
 \frac{4\pi}{2\lambda+1} \sum_{\mu=-\lambda}^\lambda Y_\lambdamu(\vartheta_i,\varphi_i) 
  Y_\lambdamu^\ast(\thetaphi) = P_\lambda(\cos\gamma_i), \\
 \cos\gamma_i \equiv \cos\vartheta \cos\vartheta_i
 +\sin\vartheta \sin\vartheta_i \cos(\varphi-\varphi_i)
\end{eqnarray*}
we obtain
\begin{eqnarray}	\label{eq:W-r} 
 W &=& \sum_{\lambda=0}^\infty \left[\frac{q_1}{m_1^\lambda}+(-1)^\lambda 
 \frac{q_2}{m_2^\lambda} \right]\left(\redmass r\right)^\lambda \sum_{i=3}^N 
 \frac{q_i}{r_i^{\lambda+1}} P_\lambda(\cos\gamma_i)	\nonumber \\
 &=& W_0(r_3,\ldots,r_N) + W_1(\pos/r,\pos_3,\ldots,\pos_N)\,r + \ldots
\end{eqnarray}
for potential energy contribution of non-coalescent particles.
\emph{For identical coalescent particles, all odd powers of separation vanish} in the vicinity of the coalescence point due to the inversion symmetry.
This property has a profound consequence in the behaviour of electron-electron potentials which will be discussed in a separate paper. 
Leading term 
\begin{equation} \label{eq:W0-def} 
 W_0 \equiv W_0\left(r_3,\ldots,r_N\right) \equiv (q_1 + q_2) \sum_{i=3}^N\frac{q_i}{r_i} 
\end{equation}
of expansion (\ref{eq:W-r}) depends on distances rather than positions of the non-coalescent particles. 
In practical calculations, the terms of above sum should be evaluated as expectation values $\matrixel{\Psi}{q_i/r_i}{\Psi}$, however the fixed-nucleus approximation leaves them in their original form for the nuclei.
Since average of Legendre polynomials with $\lambda > 0$ vanish the spherical average of $W$ equals the leading term of its expansion
\begin{equation}	\label{eq:W-avg} 
 \overline{W} = W_0 
\end{equation}
within the convergence radius of expansion (\ref{eq:W-r}).
Consequences of this property will be discussed in detail in section \ref{subsec:Psi-0-avg}.
Figure \ref{fig:W-chain} illustrates the behaviour of potential $W$ in the vicinity of a Coulomb singularity.

Since term $\hat{W}$ of Hamiltonian (\ref{eq:hamiltonian-0}) acts on separation $\pos$ of particles 1 and 2 and term $\hat{G}$ acts on their center of mass $\Pos$ the many-particle Schr\"odinger equation is separable resulting in an effective one-body problem.
The wave equation for $\Psi = \Psi\left(\pos\vert\pos_3,\ldots,\pos_N\right)$ approaches  homogeneous differential equation
\begin{equation*}
 \left[-\frac{\Laplacian}{2\redmass} + \frac{q_1 q_2}{r} + W_0 - E + \Ordo{\pos} \right]\Psi
 \approach{r}{0} 0  
\end{equation*}
in the vicinity of coalescence points, where $E$ is an eigenvalue of many-electron Schr\"odinger equation (\ref{eq:schrodinger}).
Due to local isotropy (\ref{eq:commut-0}) of Hamiltonian (\ref{eq:hamiltonian-0}) the equation is locally separable in terms of spherical polar coordinates and the wave function exhibits local hydrogenic angular dependence of the form (\ref{eq:Psi-hyd-0}) about the origin.
The corresponding radial wave equation 
\begin{equation*}
 \left\lbrace \frac{-1}{2\redmass r^2}\left[ \deriv{}{r}{} \left( r^2 \deriv{}{r}{} \right) 
 - \ell(\ell+1) \right] 
 + \frac{q_1 q_2}{r} + W_0 - E + \Ordo{\pos} \right\rbrace R \approach{r}{0} 0  
\end{equation*}
has an \emph{isolated} singular point at the origin. 
Frobenius normal form of this equation is
\[  r^2 R^{\prime\prime} + r P(r) R^\prime + Q(r)R = 0 , \]
where
\begin{equation*}
 P \equiv 2, \quad
 Q \equiv - \ell(\ell+1) - 2\redmass q_1 q_2 r - 2\redmass(W_0 - E)r^2 + \Ordo{\pos^3}. 
\end{equation*}
The singular point is \emph{removable} since $P(r)$ and $Q(r)$ are single-valued analytic functions.
\emph{Fuchs' theorem} \cite{Schlesinger1900, Ince1944, Whittaker1963} states that in the neighbourhood of removable singularities, the fundamental system of solutions is
\begin{eqnarray*}
 R_1 &=& r^{\lambda_1} u(r),	\\
 R_2 &=& r^{\lambda_2} \left[ v(r) + \alpha \, u(r) \ln r \right],
\end{eqnarray*}
where $\Re \lambda_1 \geq \Re \lambda_2$, $u(r), \: v(r)$ are single-valued analytic functions, $u(0) \neq 0$, $v(0) \neq 0$ and $\alpha$ is a constant 
\footnote{Transformation of $r$ to a dimensionless variable required by the logarithmic term is omitted for brevity.}.
The theorem distinguishes three cases for existence of logarithmic term of the second solution depending on difference $\lambda_1 - \lambda_2$.
In order to determine exponents $\lambda_1$ and $\lambda_2$ we seek the solution in the form $r^\lambda u(r)$. 
The substitution gives \emph{indicial equation}
\begin{equation*}
 \left[\lambda(\lambda-1) + 2\lambda - \ell(\ell+1)\right]u(r) + \Ordo{r} 
 \approach{r}{0} 0 
\end{equation*}
of which roots are $\lambda_1 = \ell$ and $\lambda_2 = -\ell-1$.
Since difference $\lambda_1-\lambda_2=2\ell+1$ is a non-zero integer there is no general rule for existence of logarithmic term hence value of $\alpha$ should be determined individually by substituting $R_2$ into the differential equation leading to 
\begin{equation*}
 \alpha \approach{r}{0} 
 - r \frac{2\redmass q_1 q_2 v(r) + 2\ell v^\prime(r) + \Ordo{r}}{(2\ell+1)u(r) + \Ordo{r}}
 \approach{r}{0} 0, 
\end{equation*}
i.e. the logarithmic term of the second solution vanishes at the origin hence the fundamental solutions are simply $R_1 = r^\ell u(r)$ and $R_2 = r^{-\ell-1} v(r)$.
Since $R_2$ is unbounded at $r=0$ the physical solution is
\begin{equation}	\label{eq:R-0} 
 R \approach{r}{0} r^\ell u(r).
\end{equation} 
Substituting this expression into the radial equation we obtain differential equation
\begin{equation}	\label{eq:Sch-u} 
 u^{\prime\prime} + \frac{2\ell+2}{r} u^\prime 
 - 2\redmass \left[ \frac{q_1 q_2}{r} + W_0 - E + \Ordo{\pos} \right] u 
 \approach{r}{0}  0 
\end{equation}
of which solution should be analytic according to Fuchs' theorem hence it can be expanded in power series 
\begin{equation}	\label{eq:u-ser} 
 u(r) = u(0) + u^\prime(0)r + \frac{u^{\prime\prime}(0)}{2!}r^2 + \ldots,
 \quad u(0) \neq 0.
\end{equation}
Inserting it into the differential equation we obtain algebraic equation
\begin{eqnarray*}
 - 2\left[\redmass q_1 q_2 u(0)-(\ell+1)u^\prime(0)\right]r^{-1} \\
 + \left[2\redmass (E-W_0)u(0)-2\redmass q_1 q_2 u^\prime(0)
 + (2\ell+3)u^{\prime\prime} (0)\right]r^0 + \Ordo{\pos} \approach{r}{0} 0 
\end{eqnarray*}
which can be satisfied only if
\begin{subequations} 
\begin{eqnarray}
 \label{eq:u-cusp1} a &\equiv& \frac{u^\prime(0)}{u(0)} = \frac{\redmass q_1 q_2}{\ell+1}, \\
 \label{eq:u-cusp2} b &\equiv& \frac{1}{2} \frac{u^{\prime\prime}(0)}{u(0)}
 = \frac{(\ell+1)a^2 + \redmass\left(W_0-E\right)}{2\ell+3},
\end{eqnarray}
\end{subequations} 
where \emph{cusp condition} (\ref{eq:u-cusp1}) removes singularity $\Ordo{r^{-1}}$ of the equation and relation (\ref{eq:u-cusp2}) represents a \emph{constraint on curvature} of the wave function at the origin \footnote{Leading expansion coefficients of radial wave functions of the $H$ atom are the same as (\ref{eq:u-cusp1}, \ref{eq:u-cusp2}) with $W_0=0$.}.
We have to note that $b=b\left(r_3,\ldots,r_N\right)$ depends only on charges and distances of the non-coalescent particles similarly to $W_0=W_0\left(r_3,\ldots,r_N\right)$.

Since singularity of many-particle Hamiltonian (\ref{eq:hamiltonian-0}) is isotropic at $r=0$ its eigenfunctions have the form
\begin{equation} \label{eq:Frobenius-ser} 
 \Psi = r^\ell u\left(\pos\vert\pos_3,\ldots,\pos_N\right)
 = r^\ell \sumlambdamu r^\lambda u_\lambdamu(r) Y_\lambdamu(\thetaphi) 
\end{equation}
in the vicinity of coalescence points which can be considered as the spatial generalization of Frobenius series 
\footnote{Due to ignoring the asymptotic central symmetry of Hamiltonian at $r=0$, spatially generalized power series $\Psi=\sumlambdamu r^\lambda u_\lambdamu(r;\pos_3,\ldots,\pos_N) Y_\lambdamu(\thetaphi)$ was used by former authors.}
, where $u_\lambdamu(r) \equiv u_\lambdamu(r\vert\pos_3,\ldots,\pos_N)$.
In view of local commutativity (\ref{eq:commut-0}) of angular momentum with Hamiltonian both $\ell$ and $m$ have definite values at the origin.
Since many-particle wave function is antisymmetric under interchange of any two electrons, $\ell$ takes only even values for relative singlet spin states $s_1+s_2=0$ and odd values for relative triplet spin states $s_1+s_2=1$ of the coalescent electrons.
Since leading terms $\ell(\ell+1)/2\redmass r^2$, $q_1 q_2/r$ and $W_0$ of Hamiltonian are isotropic, three leading terms of above expansion exhibit hydrogenic angular dependence 
\begin{equation*}
 \Psi\approach{r}{0} r^\ell u_\ell(0)\left(1+ar+br^2\right) Y_\lm(\thetaphi)+\ldots\,. 
\end{equation*}
Therefore the many-electron wave function can be decomposed in the vicinity of Coulomb singularities as
\begin{equation}	\label{eq:Psi-ser-0}
 \Psi = r^\ell \left[ u_\ell ( r\vert\pos_3,\ldots,\pos_N) Y_\lm(\thetaphi)
                    + r^3  v(\pos\vert\pos_3,\ldots,\pos_N) \right] ,
\end{equation}
where
\begin{eqnarray*}
 u_\ell &=& u_\ell\left(0\right)\left(1+ar+br^2+cr^3+\ldots\right), \\
 v &=& \sum_{\lambda=\ell+1}^\infty \sum_{\mu=-\lambda}^\lambda
 r^{\lambda-\ell-1} v_{\lambda\mu}(r\vert\pos_3,\ldots,\pos_N) Y_\lambdamu\left(\thetaphi\right).
\end{eqnarray*}
Term $\Ordo{\pos^\ell}$ of decomposition (\ref{eq:Psi-ser-0}) is an eigenfunction of the spherically averaged Hamiltonian, term $\Ordo{\pos^{\ell+3}}$ reflects anisotropy of molecular or crystalline potential, so group theoretical considerations apply only to $v=v(\pos\vert\pos_3,\ldots,\pos_N)$.


\subsection{Local behaviour at Coulomb singularities of spherically averaged Hamiltonian \label{subsec:Psi-0-avg}}
Let us investigate the local behaviour of first term of decomposition (\ref{eq:Psi-ser-0}) in more detail.
In view of Eq. (\ref{eq:W-avg}) the limiting form of spherically averaged Hamiltonian is equivalent to that of a Coulomb potential embedded in a uniform background
\begin{equation}	\label{eq:H-0-avg} 
 \overline{\hat{H}} \approach{r}{0} - \frac{\Laplacian}{2\redmass} + \frac{q_1 q_2}{r} + W_0
\end{equation}
hence Eq. (\ref{eq:Sch-u}) can be rewritten as
\[ ru^{\prime\prime}+\left(2\ell+2\right)u^\prime-\left(2\alpha+\beta^2 r\right)u=0,\]
where $\alpha \equiv \redmass q_1 q_2$ and $\beta^2 \equiv 2\redmass\left(W_0-E\right)$.
Seeking the solution in the form $u=\rme^{-\beta r}w(r)$ and then by making change of variable $x \equiv 2\beta r$ we obtain following confluent hypergeometric equation 
\begin{equation*}
 xw^{\prime\prime}+\left(2\ell+2-x\right)w^\prime-\left(\ell+1+\alpha/\beta\right)w=0
\end{equation*}
which is of Kummer type \cite{Abramowitz1972}
\begin{equation*}
 xw^{\prime\prime}+\left(\mathfrak{b}-x\right)w^\prime-\mathfrak{a}w = 0
\end{equation*}
with $\mathfrak{a}=\ell+1+\alpha/\beta$ and $\mathfrak{b}=2\ell+2$.
Regular solution of this equation is the following Kummer function 
\footnote{The solution reduces to associated Laguerre polynomial in case of $W_0=0$. We may formally consider $-\alpha/\beta$ as a non-integer principal quantum number.}
\begin{equation*}
 w={_1}F_1\left(\mathfrak{a}; \, \mathfrak{b}; \, x\right) \equiv \sum_{k=0}^\infty
  \frac{\left(\mathfrak{a}\right)_k}{\left(\mathfrak{b}\right)_k} \frac{ x^k}{k!}, 
\end{equation*}
where $\left(\mathfrak{a}\right)_k$ denotes the Pochhammer symbol defined by 
$\left(\mathfrak{a}\right)_k \equiv \Gamma\left(\mathfrak{a}+k\right) / \Gamma\left(\mathfrak{a}\right)$.
Therefore the physical solution of Eq. (\ref{eq:Sch-u}) in the neighbourhood of $r=0$ is
\begin{equation}	\label{eq:u-0-avg} 
 u_\ell(r)=\rme^{-\beta r} {_1}F_1\left(\ell+1+\alpha/\beta;\,2\ell+2;\,2\beta r\right).
\end{equation}

Since Eq. (\ref{eq:Sch-u}) is non-singular, its analytic solution can be expanded in terms of a power series.
By substituting power series $u_\ell(r)=\sum_{k=0}^\infty a_k r^k$ into the equation and 
collecting terms with equal powers of $r$ we arrive at algebraic equation
\begin{eqnarray*}
   && \left[ \left(2\ell+2\right)a_1 - 2\alpha a_0 \right] r^0 \\
 + && \sum_{k=1}^\infty\left[(k+1)(2\ell+2+k)a_{k+1}-2\alpha a_k-\beta^2 a_{k-1}\right] r^k = 0
\end{eqnarray*}
leading to recurrence relations
\begin{subequations}
\begin{eqnarray}
 \label{eq:recurrence-1}
 a_1 &=& \frac{\alpha}{\ell+1} a_0,  \\
 \label{eq:recurrence-k}
 a_{k+1} &=& \frac{2\alpha \, a_k + \beta^2 a_{k-1}}{(2\ell+2+k)(k+1)},
 \quad k=1,2,\ldots \, .	  
\end{eqnarray}
\end{subequations}
The first one is equivalent to  L\"owdin's \cite{Lowdin1954} first order cusp condition (\ref{eq:lowdin}) and the second one provides a simple way of generating higher order cusp conditions for eigenfunctions of the spherically averaged Hamiltonian (\ref{eq:H-0-avg}).
Use of these recurrence relations is more efficient numerically than evaluating the confluent hypergeometric function in Eq. (\ref{eq:u-0-avg}).

\begin{figure} 
\includegraphics[width=4.5in]{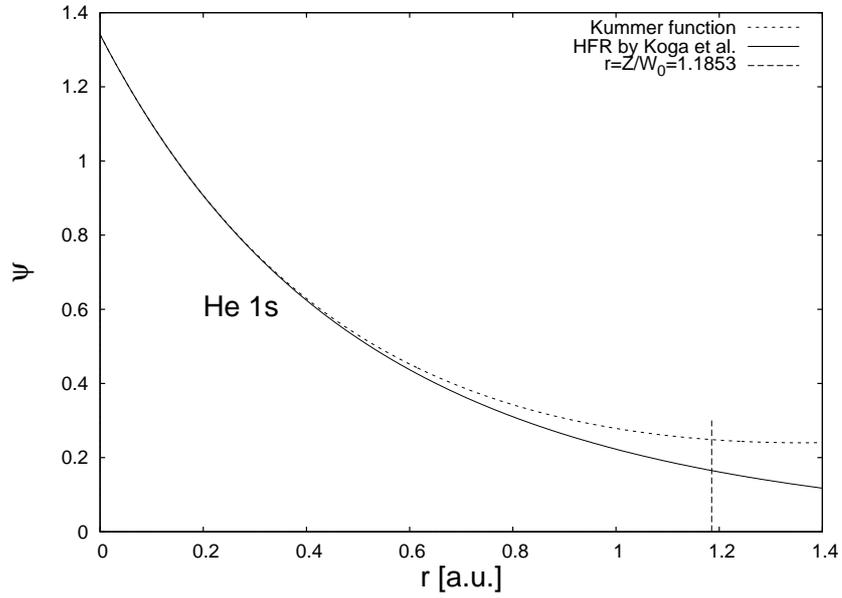}
\caption{\label{fig:he-kummer}
Wave function of Kummer type (\ref{eq:psi-0-avg}) compared with Hartree-Fock-Roothaan (HFR) wave function computed by Koga, Kanayama, Watanabe and Thakkar \cite{Koga1999} for ground state of the $He$ atom. Radius $r=Z/W_0$ of bound-state region of spherically averaged Hamiltonian (\ref{eq:H-0-avg}) is marked on the $r$ axis.}
\end{figure}

\begin{figure} 
\includegraphics[width=4.5in]{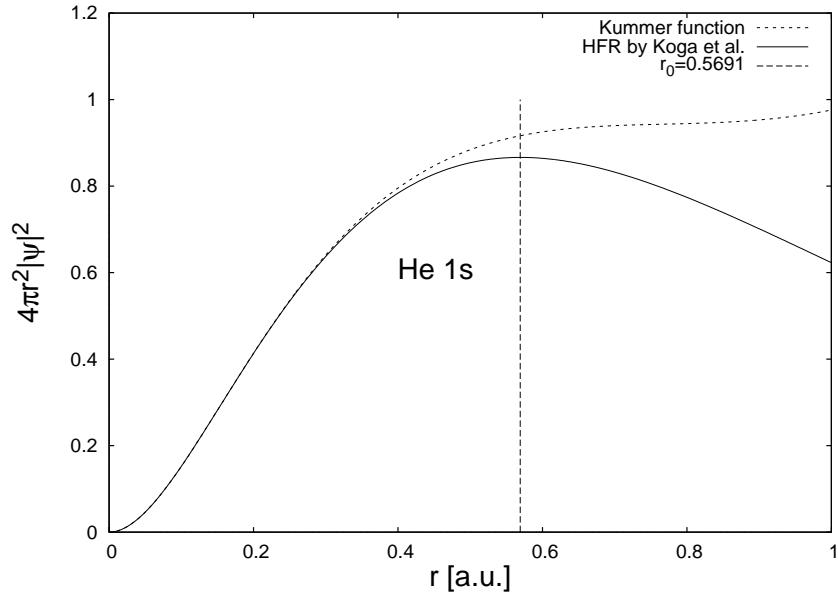} 
\caption{\label{fig:he-kummer-den}
Radial density obtained from (\ref{eq:psi-0-avg}) compared with that of computed by Koga et al. \cite{Koga1999} for ground state of the $He$ atom. 
Our near-nucleus approximation is accurate up to surprisingly large distances.
Effective Bohr radius $r_0$ of the orbital is marked on the $r$ axis.}
\end{figure}

In view of Eqs. (\ref{eq:Psi-hyd-0}), (\ref{eq:R-0}) and (\ref{eq:u-0-avg}) the eigenfunction of the spherically averaged Hamiltonian is of the form
\begin{equation}	\label{eq:psi-0-avg} 
 \psi =  r^\ell u_\ell(0) \rme^{-\beta r} 
 {_1}F_1\left(\ell+1+\alpha/\beta;\, 2\ell+2;\, 2\beta r\right) Y_\lm(\vartheta,\varphi)
\end{equation}
in the vicinity of Coulomb singularities (for comparison with an accurate Hartree-Fock-Roothaan atomic wave function see Figures \ref{fig:he-kummer} and \ref{fig:he-kummer-den}).
This analytic function fully characterizes cusps of spherically symmetric or spherically averaged systems since arbitrarily high order cusp relations can be derived from it. 
Higher order cusp conditions obtained from Kummer type wave function (\ref{eq:psi-0-avg}) agree with that of Refs. \cite{Rassolov1996} and \cite{Nagy2001} (after typographic correction $Z_\alpha \mapsto Z_\alpha^2$).
For anisotropic systems, only first and second order cusp conditions (\ref{eq:u-cusp1}, \ref{eq:u-cusp2}) are exact which can also be obtained from recurrence relations (\ref{eq:recurrence-1}, \ref{eq:recurrence-k}).
This bound-state local solution becomes non-physical at larger distances satisfying $q_1 q_2/r+W_0 \geq 0$ since non-coalescent particles within the sphere of radius $r$ are not included.
Function (\ref{eq:psi-0-avg}) is not square-integrable but knowledge of its higher order derivatives at $r=0$ is useful in numerical calculation of first term of Eq. (\ref{eq:Psi-ser-0}).

\section{Many-electron cusp and boundary conditions \label{sec:cusp-bc}}
Using expansions (\ref{eq:Psi-ser-inf}) and (\ref{eq:Psi-ser-0}) of many-electron wave function about singular points of Hamiltonian (\ref{eq:hamiltonian}) we are able to recover explicit forms of natural boundary conditions for the many-electron Schr\"odinger equation.
Due to local isotropy (\ref{eq:commut-0}, \ref{eq:commut-inf}) of Hamiltonian at singular points we can follow the same procedure as we did in section \ref{sec:bc-qm} for the $H$ atom.
One of our resulted boundary conditions may be considered as an \emph{exact} cusp condition without spherical average which contradicts Kato's result.
Using our Kummer function solution (\ref{eq:psi-0-avg}) for the spherically averaged Hamiltonian we can define arbitrarily high order cusp conditions in the framework of central field approximation.

\subsection{Boundary conditions \label{subsec:bc} }
In view of expansion (\ref{eq:Psi-ser-inf}) of many-electron wave function at large distances, its logarithmic derivative with respect to $r$ is
\begin{equation*}
 \frac{1}{\Psi} \frac{\partial \Psi}{\partial r} \approach{r}{\infty} 
 -\sqrt{-2\redmass^\prime  E} + \left[\frac{\redmass^\prime(Q+1)}{\sqrt{-2\redmass^\prime E}} - 1\right]
 \frac{1}{r} + \Ordo{\pos^{-2}} 
\end{equation*}
which defines our first boundary condition:
\begin{equation}	\label{eq:bc-Psi-inf} 
 \lim_{r \to \infty} \frac{1}{\Psi} \frac{\partial \Psi}{\partial r} 
 = -\sqrt{-2\redmass^\prime E}.
\end{equation}

The directional logarithmic derivative of the many-electron wave function in arbitrary direction $\unitvec$ can be expressed in terms of spherical polar coordinates as
\begin{equation*}
 \frac{\unitvec\cdot\nabla\Psi}{\Psi} 
 = \frac{\unitvec\cdot\unitvecr}{\Psi} \frac{\partial\Psi}{\partial r}
 + \frac{\unitvec\cdot\unitvectheta}{\Psi\,r} \frac{\partial\Psi}{\partial\vartheta}
 + \frac{\unitvec\cdot\unitvecphi}{\Psi\,r\sin\vartheta} 
   \frac{\partial\Psi}{\partial\varphi}, 
\end{equation*}
where $\polarbasis$ is basis of the coordinate system.
In radial directions, defined by $\unitvec\cdot\unitvecr=\pm1$ and $\unitvec\cdot\unitvectheta=\unitvec\cdot\unitvecphi=0$, above expression reduces to
\begin{equation*}
 \frac{\unitvec\cdot\nabla\Psi}{\Psi}
 = \frac{\unitvec\cdot\nabla_r\Psi}{\Psi}
 = \frac{\pm 1}{\Psi} \frac{\partial\Psi}{\partial r}. 
\end{equation*}
Since 
wave function $\Psi=r^\ell u(\pos\vert\pos_3,\ldots,\pos_N) \equiv r^\ell u(\pos)$ given by Eq.  (\ref{eq:Frobenius-ser}) has a root of multiplicity $\ell$ at Coulomb singularities the l'Hospital rule should be applied $\ell$ times in order to obtain a definite limit
\begin{equation*}
 \frac{1}{\Psi} \frac{\partial\Psi}{\partial r} \, \approach{r}{0} \,
 \frac{\partial_r^{\ell+1} \Psi}{\partial_r^\ell \Psi} \, \approach{r}{0} \,
 \frac{\ell+1}{u(\pos)} \frac{\partial u(\pos)}{\partial r} + \Ordo{\pos},
\end{equation*}
where differentiation rules (\ref{eq:der-rad-l}, \ref{eq:der-rad-l+1}) are used
\footnote{In general, radial partial derivative of an anisotropic function is anisotropic which explains appearance of spherical average in Kato's cusp condition (\ref{eq:kato}). See discussion after the statement of theorem $IIa$ in Ref. \cite{Kato1957}.}.
Since expansion (\ref{eq:Psi-ser-0}) exhibits hydrogenic angular dependence $r^\ell u_\ell(r) Y_\lm(\vartheta,\varphi)$ in the vicinity of Coulomb singularities the radial logarithmic derivative of $\Psi$ has the following one-sided limits:
\begin{equation}	\label{eq:bc-Psi-0}
 \lim_{\pos \to \pm 0} \frac{1}{\Psi} \frac{\partial\Psi}{\partial r} 
 = \lim_{\pos \to \pm 0} \frac{\partial_r^{\ell+1} \Psi}{\partial_r^\ell \Psi}
 = \pm \left(\ell+1\right) a,
\end{equation}
where $a$ is defined by Eq. (\ref{eq:u-cusp1})
and
\begin{eqnarray*}
 \mathbf{r} \to +0 &\equiv& (r \to 0, \,\vartheta,     \,\varphi),		\\
 \mathbf{r} \to -0 &\equiv& (r \to 0, \,\pi-\vartheta, \,\pi+\varphi).
\end{eqnarray*}
Equation (\ref{eq:bc-Psi-0}) is our second boundary condition.

Equations (\ref{eq:bc-Psi-inf}) and (\ref{eq:bc-Psi-0}) represent homogeneous Robin boundary conditions for the many-electron wave function and have the same form as Eqs. (\ref{eq:bc-inf}) and (\ref{eq:bc-nuc}) obtained for the $H$ atom which is a consequence of asymptotic isotropy (\ref{eq:commut-0}, \ref{eq:commut-inf}) of Hamiltonian at singular points.

\subsection{Cusp conditions \label{subsec:cusp} }
Boundary condition (\ref{eq:bc-Psi-0}) is the exact form of first order cusp condition which does not contain spherical average.
By substituting specific values of $\redmass$, $q_1$ and $q_2$ into Eq. (\ref{eq:bc-Psi-0}) we obtain cusp conditions for the electron-nucleus coalescence
\begin{equation}  \label{eq:Psi-cusp-nuc} 
 \lim_{\pos \to \pm 0} \frac{\partial_r^{\ell+1} \Psi}{\partial_r^\ell \Psi}
 = \mp \frac{M_\nu}{M_\nu + 1} Z_\nu, \quad \ell=0,1,2,\ldots,
\end{equation}
where $M_\nu=Z_\nu m_p/m_e + \left(A_\nu - Z_\nu\right)m_n/m_e$ denotes mass of the $\nu$-th nucleus (in a. u.).
Similarly we obtain following cusp conditions for electron-electron coalescences
\begin{subequations}
\begin{eqnarray}
 &&\lim_{\pos \to \pm 0} 
 \frac{\partial_r^{\ell+1} \Psi_\spinsinglet}{\partial_r^\ell \Psi_\spinsinglet}
 = \pm \frac{1}{2}, \quad \ell=0,2,4,\ldots, \label{eq:Psi-cusp-el-sing} \\
 &&\lim_{\pos \to \pm 0} 
 \frac{\partial_r^{\ell+1} \Psi_\spintriplet}{\partial_r^\ell \Psi_\spintriplet}
 = \pm \frac{1}{2}, \quad \ell=1,3,5,\ldots, \label{eq:Psi-cusp-el-trip}
\end{eqnarray}
\end{subequations}
where $\spinsinglet$ and $\spintriplet$ denote relative singlet and triplet spin states of coalescent electrons, respectively.
First order many-body cusp conditions (\ref{eq:Psi-cusp-nuc}, \ref{eq:Psi-cusp-el-sing}, \ref{eq:Psi-cusp-el-trip}) are obviously $CP$-invariant similarly to Eq. (\ref{eq:bc-nuc}) obtained for the $H$ atom.

Since second directional derivative $\left(\unitvec\cdot\nabla\right)^2$ is of definite sign the second order cusp condition has the form
\begin{equation}	\label{eq:Psi-cusp2} 
 \lim_{r \to 0} \frac{1}{\Psi} \frac{\partial^2 \Psi}{\partial r^2}
 = \lim_{r \to 0} \frac{\partial_r^{\ell+2}\Psi}{\partial_r^\ell \Psi}
 = \left(\ell+1\right)\left(\ell+2\right) b, 
\end{equation}
where differentiation rules (\ref{eq:der-rad-l}, \ref{eq:der-rad-l+2}) are used, $b$ is defined by Eq. (\ref{eq:u-cusp2}).

Coefficient $a$ in Eqs. (\ref{eq:bc-Psi-0}, \ref{eq:Psi-cusp2}) depends only on charges and masses of the coalescent particles.
In view of (\ref{eq:W0-def}) and (\ref{eq:u-cusp2}), coefficient $b=b\left(r_3,\ldots,r_N\right)$ in Eq. (\ref{eq:Psi-cusp2}) depends only on charges and distances of non-coalescent particles hence it has the same value for the spherically averaged Hamiltonian.
Arbitrarily high order cusp conditions can be obtained from Eq. (\ref{eq:psi-0-avg}) by means of recurrence relation (\ref{eq:recurrence-1}, \ref{eq:recurrence-k}) in the framework of central field approximation.
Third and higher order cusp conditions for exact wave functions require spherical average similarly to Kato's cusp condition (\ref{eq:kato}) since higher order coefficients of expansion (\ref{eq:Psi-ser-0}) depend both on distances and directions of the non-coalescent particles.

In view of Eq. (\ref{eq:Psi-ser-0}) the electron pair density exhibits short-range hydrogenic angular dependence in the vicinity of coalescence points hence satisfies exact first and second order cusp conditions similar to (\ref{eq:Psi-cusp-el-sing}, \ref{eq:Psi-cusp-el-trip}) and (\ref{eq:Psi-cusp2}).

\section{Consequences \label{sec:consequences}}

\subsection{Physical consequences}
Since jump of logarithmic derivative of wave function is $2Z$ at the nuclei both in the one-electron and in the many-electron cases, its discontinuity is caused by common terms of the two Hamiltonians, namely, by singular Coulomb potential and by singular kinetic energy of opposite sign as if no other particles were present except the coalescent ones. 
An electron-electron potential taking part in cancellation of the nuclear Coulomb singularity would require a more singular wave function than the square-integrable functions hence cusp condition shows an evidence for the normalization condition.

Potential energy term $W_0$ defined by Eq. (\ref{eq:W0-def}) is related to the NMR chemical shift since it is proportional to ratio $H_i(0)/H$ of induced diamagnetic shielding field to the applied external magnetic field \cite{Lamb1941, Dickinson1950}. 


Expansion (\ref{eq:Psi-ser-0}) characterizes the short-range behaviour of many-electron 
wave function both in the vicinity of electron-nucleus and electron-electron coalescences hence it can be used to describe short-range electron correlation which depends on the chemical environment in second order through parameter $b=b\left(r_3,\ldots,r_N\right)$. 

First order cusp condition (\ref{eq:u-cusp1}, \ref{eq:bc-Psi-0}) is responsible for boundedness of many-electron wave function at Coulomb singularities by exactly cancelling singular kinetic and potential energy terms.
Second order cusp conditions (\ref{eq:u-cusp2}, \ref{eq:Psi-cusp2}) enforce correct value $E$ of local energy $E_{loc}\equiv\Psi^{-1}\hat{H}\Psi$ at Coulomb singularities by imposing constraints on curvature of many-electron wave function.
These conditions are exact for arbitrary values of $\ell$ and do not need spherical average in contrast to Kato's cusp condition (\ref{eq:kato}).

One can see from Eq. (\ref{eq:psi-0-avg}) that all radial partial derivatives of eigenfunctions of spherically averaged Hamiltonian at coalescence points depend on the same constant quantities: $\ell$, $q_1$, $q_2$, $M$, $E$ and $W_0$.

Third and higher derivatives of exact eigenfunctions depend on anisotropic multipole potential terms of expansion (\ref{eq:W-r}) as well.
In view of Eq. (\ref{eq:Psi-ser-0}), symmetry considerations apply only to terms $\Ordo{\pos^{\ell+3}}$. 

The real component of Bloch functions exhibits cusps at the nuclei similarly to molecular wave functions since the cosine function does not vanish at $r=0$.

\subsection{Numerical consequences}
If both boundary conditions are known the numerical solution of Schr\"{o}dinger equation over a grid yields an algebraic eigenvalue problem 
which can be solved by means of Jacobi - Goldstine - Murray - von Neumann diagonalization algorithm \cite{Goldstine1959}.
Before Kato's cusp condition only long-range behaviour (\ref{eq:bc-Psi-inf}) of wave function was known hence a numerical trick called \emph{shooting method} was used to substitute the missing boundary condition at Coulomb singularities.
This method is best suited to equidistant grids, where inward and outward numerical integrations are equally accurate.
Since most of energy is concentrated at near-nucleus regions the practical grids used in quantum chemistry are substantially finer in the vicinity of nuclei than in the interstitial and exterior regions.
The grid is coarsest at large distances representing $\infty$ in the numerical calculation hence an outward integrated solution starting from a guessed initial condition is fitted at a midpoint to an inward integrated solution based on an inaccurate initial condition.
Kato's cusp condition (\ref{eq:kato}) is suitable only to the central-field approximation and cannot be used as an exact boundary condition.
Our boundary conditions (\ref{eq:bc-Psi-inf}) and (\ref{eq:bc-Psi-0}) are exact regardless of molecular or crystalline symmetry since the wave function exhibits hydrogenic angular dependence in both limiting cases.
A shooting method based on our cusp condition (\ref{eq:bc-Psi-0}) would be more accurate than the traditional one due to the finer grid at nuclei.
Since both boundary conditions are known one can transform Schr\"{o}dinger equation to an algebraic eigenvalue problem instead of performing the time-consuming shooting loop.
Approximate wave function (\ref{eq:psi-0-avg}) can be used to find optimal near-nucleus step size for the grid.

Basis sets satisfying cusp conditions improve convergence of Hartree-Fock-Roothaan variational calculations.
It is well known that Slater-type basis sets satisfy first order cusp condition if exponential decay parameter of one basis function is appropriately fixed
\begin{equation*}
 R = c_0 r^\ell \rme^{-\frac{Zr}{\ell+1}} + \sum_{\lambda = 1}^L 
     c_\lambda r^{\ell+\lambda} e^{-\zeta_\lambda r} 
\end{equation*}
letting only its weight factor $c_0$ to be varied.
Similar basis set was used by Roothaan and Kelly \cite{Roothaan1963} however their summation inexplicably starts from $\lambda=3$ resulting in a slow convergence for $\ell > 0$ hence use of cusp condition was limited to the $s$ orbitals in their atomic calculations \footnote{We suppose that their summation starting from $\lambda=3$ arises from some unpublished expansion similar to our Eq. (\ref{eq:Psi-ser-0}), where our $u_\ell$ was simply replaced by $e^{-Zr/(\ell+1)}$.}.
It is easy to construct a Slater-type basis set satisfying both first and second order cusp conditions by equating three leading expansion coefficients of linear combination of two Slater functions with equal exponential factors to that of expansion (\ref{eq:u-ser}) yielding
\begin{equation} \label{eq:Slater-basis} 
 R = c_0 r^\ell \left[ 1 + \left( b-\frac{a^2}{2} \right) r^2 \right] \rme^{ar} 
   + \sum_{\lambda=3}^L c_\lambda r^{\ell+\lambda} \rme^{-\zeta_\lambda r}, 
\end{equation}
where $a$ and $b$ are defined by Eqs. (\ref{eq:u-cusp1}, \ref{eq:u-cusp2}).
In order to preserve asymptotic behaviour of wave function as $r \to \infty$ a Slater function of the form  (\ref{eq:R-inf}) should be added to the basis set.

It is widely believed that Gaussian basis sets are not suitable to describe nuclear cusps since they have zero gradients at the nuclei which is true only for individual Gauss functions but not for their linear combinations 
By equating three leading expansion coefficients of linear combination of three Gauss functions with equal exponential factors to that of expansion (\ref{eq:u-ser}) we obtain Gaussian basis set
\begin{eqnarray} \label{eq:Gauss-basis} 
 \psi(x,y,z) &=& c_0 r^\ell \left[ 1 + ar + \left( b+g_0 \right) r^2 \right] 
                 \rme^{-g_0 r^2} Y_\lm(\thetaphi)	\nonumber\\
             &&+ \sum_{i+j+k \geq \ell+3}^L c_{ijk} x^i y^j z^k \rme^{-g_{ijk}r^2}
\end{eqnarray}
satisfying both first and second order cusp conditions, where $a$ and $b$ are defined by (\ref{eq:u-cusp1}, \ref{eq:u-cusp2}) and $r^2=x^2+y^2+z^2$.
There is no finite linear combination of Gauss functions which exhibits asymptotic behaviour (\ref{eq:R-inf}) of wave function as $r \to \infty$.
Use of basis functions satisfying both first and second order electron-electron cusp conditions provides the simplest way to include short-range correlation effects.
Above basis set is more efficient numerically and requires less modification of existing Gaussian computer codes than implementing a Jastrow-type correlation \cite{Jastrow1955}.

Asymptotic hydrogenic angular dependence of three leading terms of expansion (\ref{eq:Psi-ser-0}) of many-electron wave function about Coulomb singularities explains the success of central-field approximation used in atomic calculations and muffin-tin approximation of solid state physics.
The Kummer-type confluent hypergeometric function (\ref{eq:psi-0-avg}) intended to characterize the short-range behaviour of eigenfunctions 
of the spherically averaged Hamiltonian is surprisingly accurate even at relatively large distances, e.g. one can see from Figure \ref{fig:he-kummer-den} that relative error of this approximation to radial density $4\pi r^2 \abs{\psi}^2$ is 5.8\% at the effective Bohr radius and is 0.4\% at its half.
Therefore the spherically averaged part of Hamiltonian is responsible for most of the effects and anisotropic terms can be considered as perturbations.

\appendix		
\section*{Appendix: Some differentiation rules \label{app:diff-rules}}
Leibniz's theorem for differentiation of products states that
\begin{equation*}
 \left[ f(x)g(x) \right]^{(n)} = \sum_{k=0}^n \binom{n}{k} f^{(n-k)}(x) g^{(k)}(x). 
\end{equation*}
For $f(x)=x^\ell$ one obtains
\begin{equation*}
 \left[ x^\ell g(x) \right]^{(n)} = \sum_{k=0}^n \binom{n}{k} 
  \frac{\ell! \, x^{\ell-n+k} g^{(k)}(x)}{(\ell-n+k)!}. 
\end{equation*}
Specific higher order derivatives of the above type used in this paper are
\begin{subequations}
\begin{eqnarray}
 \label{eq:der-l} \left[ x^\ell g(x) \right]^{(\ell)} 
 &&\approach{x}{0} \ell! \, g(0) + \Ordo{x},	\\
 \label{eq:der-l+1} \left[ x^\ell g(x) \right]^{(\ell+1)} 
 &&\approach{x}{0} (\ell+1)! \, g^\prime(0) + \Ordo{x}, \\
 \label{eq:der-l+2} \left[ x^\ell g(x) \right]^{(\ell+2)} 
 &&\approach{x}{0} (\ell+2)! \, g^{\prime\prime}(0) + \Ordo{x}.
\end{eqnarray}
\end{subequations}
For higher order radial partial derivatives of $r^\ell u(\pos)$ we obtain similarly
\begin{subequations}
\begin{eqnarray}
 \label{eq:der-rad-l}  \frac{\partial^\ell r^\ell u(\pos) }{\partial r^\ell}
 &&\approach{r}{0} \ell! \, u(0) + \Ordo{\pos}, \\
 \label{eq:der-rad-l+1}  \frac{\partial^{\ell+1} r^\ell u(\pos)}{\partial r^{\ell+1}}
 &&\approach{r}{0} (\ell+1)! \, \at{\frac{\partial u}{\partial r}}{r=0}+\Ordo{\pos}, \\
 \label{eq:der-rad-l+2}  \frac{\partial^{\ell+2} r^\ell u(\pos)}{\partial r^{\ell+2}}
 &&\approach{r}{0} (\ell+2)! \, \at{\frac{\partial^2 u}{\partial r^2}}{r=0} + \Ordo{\pos}. 
\end{eqnarray}
\end{subequations}
Quotients of above derivatives become isotropic if $u(\pos)$ can be written as a product of radial and angular parts.

\bibliography{bc_many_el}	
\end{document}